# Useful vacancies in Single Wall Carbon Nanotubes

A. Proykova[a], H. Iliev[a], and Feng Yin Li[b]

[a]University of Sofia, Faculty of Physics, 5 James Bourchier Blvd. Sofia-1126, Bulgaria
[b] National Chung Hsing University, Taiwan

The electronic and structural properties of zigzag and armchair single-wall carbon nanotubes (SWCNT) with a single vacancy or two vacancies located at various distances have been obtained within the frame of the Density Function Theory (DFT) and a Molecular Dynamics method. It is found that the vacancy defects interact at long ranges in armchair SWCNTs unlike the short-range interaction in zigzag SWCNTs. The density of states for different vacancy densities shows that the local energy gap shrinks with the vacancy density increase. This and other results of the investigation provide insight into understanding the relation between the local deformation of a defective nanotube and its measurable electronic properties.

**Introduction**

The design and reliability of carbon-nanotube-circuits have become a hot research area since it has been found that a carbon nanotube (CNT) is a working transistor (1). The CNTs are usually assumed as perfect crystalline wires in most models developed to describe their unique properties. However, even high-quality Single-Wall Carbon Nano Tubes (SWCNTs) on average contain one structural defect per 4 μm (2). That is why it is important to find out the influence of defects on the property of interest. Among the structural defects in SWCNTs vacancy defects are particularly important as they could be introduced in a controllable way into nanotubes aplying ionic or electronic irradiation (3).

We have previously demonstrated that defective SWCNTs posses interesting mechanical (4) and electric (5) properties that open new potential applications in nanodevices. Molecular dynamics simulations revealed that it is the vacancy distribution playing the major role in a (10,0) SWCNT resistance to external mechanical forces (4). Combined quantum mechanical and molecular mechanical calculations demonstrated that the energy gap of a (10,0) SWCNT reduces if 1% vacancies are presented (5). Biel *et al.* examined the characteristics of the localization regime in terms of the length, temperature, and density of defects of SWCNTs by averaging over various random configurations of defects for a metallic (10,10) SWCNT (6). The results of selected electrochemical deposition show that most of the electronic features of a particular CNT transistor are due to vacancy defects (2). Kim *et al.* explained the modified electrical characteristics of SWCNT with the localized gap states, found far from the band gap edge, produced by vacancy defects (7). These localized gap states, also called 'deep levels', can be spatially resolved by the scanning tunneling spectroscopy (8). The energy gap between the deep levels and the conduction or valence band can be measured by scanning photoluminescence microscopy. A model to explain the dependence of the electrical conductivity on the vacancy density is proposed by Baskin et al (9).

While it is already accepted that vacancies have measurable effects on the electrical and mechanical properties, the mechanism is not well understood in the sense that different scenarios emphasize fundamentally different origins.

The investigation reported in this short article answers to the following questions: what is the change of the bandstructure of a SWCNT after one or two defects per unit length have been introduced; what is the role of chirality in the defect-defect interactions; what is the shape of the SWCNT in the vicinity of defect(s)? Both Molecular Dynamics and *ab initio* calculations have been undertaken in the research.

These answers are needed for applications of SWCNT in nanoelectornics and, which is more important, in fundamental understanding the behavior of these low dimensional structures.

## Model SWCNTs and Computational Details

The CASTEP code has been employed to calculate the band gaps of armchair (5,5) and zigzag (10.0) SWCNTS containing 79, 159, 239 atoms with one or two vacancy defects introduced in the structure. The unit cell contains 40 atoms for the case of zigzag SWCNTs and 20 atoms for the armchair SWCNTs. A simulated nanotube is being placed in a supercell with lattice constants $a = b = 20$ Å; the lattice constant $c$ along the tube axis has been taken to be equal to the one-dimensional lattice parameter of nanotubes. The generalized gradient approximation (GGA) has been implemented for geometry optimization (10,11).

The structure of the defective nanotube is considered to be fully optimized when the force on each atom during relaxation is under 0.005 eV Å$^{-1}$. The nuclei and core electrons are represented by ultrasoft pseudopotentials (12). The summation has been performed over the 1D Brillouin zone with wavevectors varying only along the tube axis, using k-point sampling and a Monkhorst-Pack grid (13). A kinetic energy cut-off of 240 eV and 12 $k$ points are used along the $z$-axis to ensure the convergence in the calculations. The fast-fourier-transform (FFT) grid was chosen according to the number of carbon atoms in the particular SWCNT. For example, the FFT grid for the zigzag 79 carbon atom nanotube is 90x90x40, and for the armchair 79 SWCNT - 90 x 45 x 90.

The effect of vacancy density on the band gap of SWCNTs has been investigated for six zigzag and five armchair SWCNTs. Defect densities of one vacancy per 2, 3, 4, 5, 6 and 7 unit cells, corresponding to 79, 119, 159, 199, 239 and 279 carbon atoms in zigzag nanotubes have been looked at. For the case of armchair SWCNTs defect densities of one vacancy per 4, 6, 8, 10 and 12 unit cells correspond to 79, 119, 159, 199 and 239 carbon atoms.

We have also employed classical Molecular Dynamics to study the shape of defective SWCNTs and to compare the results obtained with the DFT method. Carbon-carbon interactions in the nanotube have been modeled with a semi-empirical potential due to Brenner (14). The CNT is fixed at one end thus simulating a tube that has grown on a fixed substrate. The other end is left free and atomic positions and velocities are advanced in time using velocity Verlet intergration technique.

## Results and Discussions

After structural optimization, each vacancy in all models becomes a so-called 5-1DB defect with two of its three adjacent dangling bonds (DB) in an ideal vacancy, recombining with each other to yield a pentagon ring with the remaining DB unchanged. The newly formed carbon-carbon bond is tilt about the tube axis. The same result is found in our Molecular Dynamics simulations, Fig.1 and in (15).

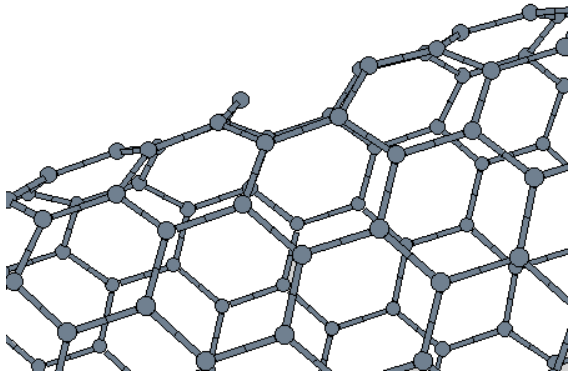
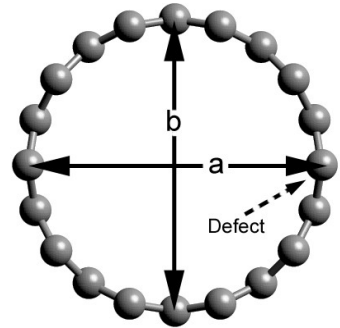

Figure 1. Molecular Dynamics optimization of a SWCNT with a vacancy: the twofold coordinated the carbon protrudes outward

Figure 2. A schematic plot to indicate the axes *a* and *b* in Table I.

The results presented in the Table I reveal that the SWCNT bulges to form an elliptical shape around the vacancy and slowly relaxes to more rounded shapes away from it. By increasing the vacancy density, the distortion of SWCNTs increases as well but the local energy gap declines. The local energy gap converges when the model includes more than 159 carbon atoms, corresponding to the distance between any two nearby vacancy defects over 16.98 Å. This sets limits on the measurable results in defective nanotubes.

Our computations show long-ranged interaction between vacancy defects in armchair SWCNTs and short-ranged interaction in zigzag nanotubes. The Table II contains simulated data for armchair nanotubes with different lengths: when the vacancy density increases, the distortion of the defected SWCNTs increases also.

**TABLE I.** Bandgap of zigzag (10,0) SWCNTs with one vacancy.

| Model Size | a(Å)* | b(Å)[&] | Local Energy Gap (eV) |
|---|---|---|---|
| 80[#] | 7.943 | 7.926 | 0.67 |
| 79 | 8.416(8.302) | 7.511(7.519) | 0.19 |
| 159 | 8.115(7.817) | 7.747(7.760) | 0.39 |
| 239 | 8.114(7.813) | 7.751(7.767) | 0.41 |

* *a* is the diameter of the ring that contains the twofold coordinated carbon; the value in parenthesis refers to the diameter of the ring equidistant between the vacancy and its image.

TABLE II. Bandgap of armchair (5,5) SWCNTs with one vacancy.

| Model Size | Local Energy Gap (eV) |
|---|---|
| 80# | 0.67 |
| 79 | 0.19 |
| 159 | 0.39 |
| 239 | 0.41 |

The strain induced by the vacancy defect can be estimated via the length variation which can be defined as the difference between the lattice constant *b* of the defective SWCNTs before and after the structural optimization. The positive length variation indicates that the lattice constant *c* expands after structural optimization. Closely examining the length variation of the defective SWCNTs with various vacancy densities, as presented in Table II reveals that formation of 5-1DB from ideal vacancy can either shrink or expand the length of a defective SWCNT as the defect density increases.

There are no obvious correlation between the vacancy density and the bandgap in the defective armchair SWCNTs. The band structures shown in the Figure 3 indicate that the valence band maximum (VBM) state and the conduction band maximum (CBM) state fluctuate dramatically among various mono-vacancy defect densities.

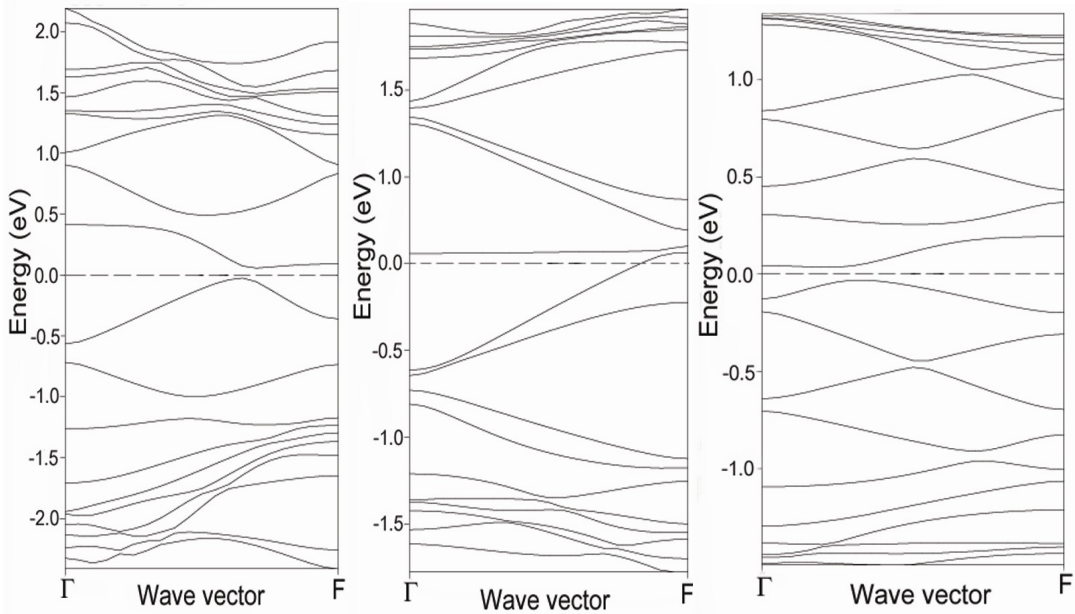

Figure 3. The bandstructures of the single mono-vacancy defected (5,5) armchair carbon nanotube with 79, 159, and 239 carbon atoms.

The density of state (DOS) for three different vacancy densities in (10,0) SWCNTs is shown in Fig. 4. Clearly the Fermi level lies between the deep levels and the VBM state, and not only the local energy gap shrinks but also the deep level becomes broaden as the vacancy density increases.

Closely examining the local structure of the 5-1DB defects with various MVD densities, it reveals that 5-1DBs shrink as the defect density increases. This shrinkage in turn causes the twofold coordinated carbon to protrude away from the tube. The same effect seen in the classical Molecular dynamics calculations is explained with the special properties of the Brenner potential which favors larger angles between the bonds connecting the carbon atoms.

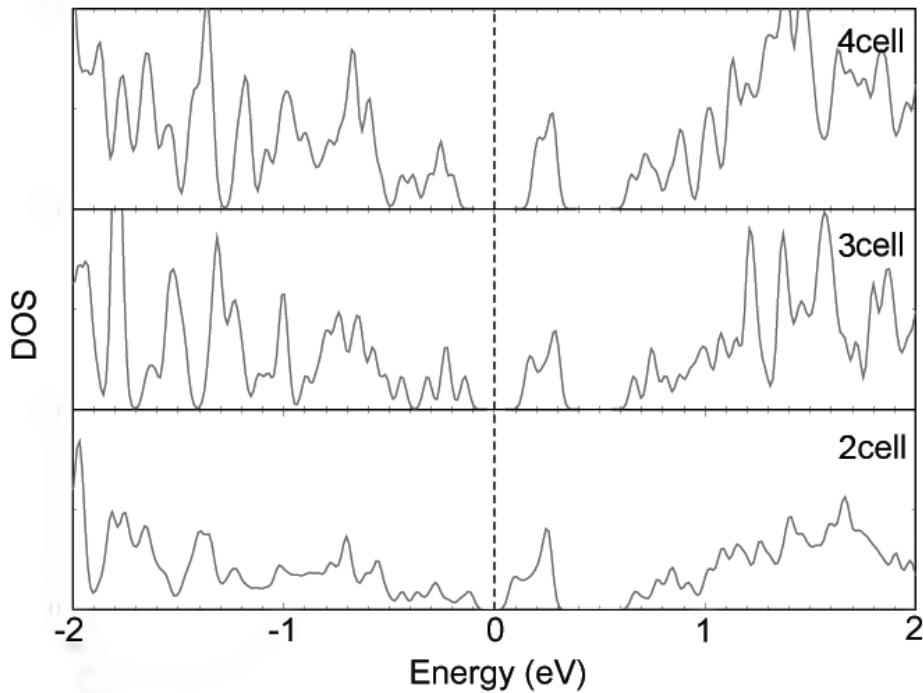

Figure 4. The DOS for three different MVD density in zigzag SWCNT. The band gaps between VBM and CBM states for 2, 3 and 4 unit cells are 0.792, 0.809 and 0.849 eV, respectively.

As to the variance of the electronic structure, the local partial DOS (LPDOS) calculation was made by analyzing those rings around the 5-1DB, presented in Fig. 5 to understand the effects of structural deformation on the electronic structure of SWCNTs, where the 2P orbitals of carbon atoms are dissected into two categories: along the tube axis (Pz) and radial components (Px and Py). Our results indicate that VBM state is composed mostly by the Px and Py orbitals of all the carbon atoms. However, at a higher vacancy density, the deep levels are composed by the Pz orbital of the twofold coordinated carbon and the Px and Py orbitals of the alternative rings, including the ring with the newly formed C-C bond. Therefore, at least in this last case, the charge density of the deep levels are not only localized in the three carbon atoms surrounding the vacancy but also include the Px and Py orbitals of the alternative rings around the vacancy site. Given the orbital distributions, the interaction between the twofold coordinated carbon and the rest of the carbon atoms substantially influence the bandstructure of the nanotube, and especially the deep levels. To put all the results together, the overall picture emerges. At a high defect density, in order to release the strain caused by surrounding defects, the 5-1DB shrinks and then causes the twofold coordinated carbon protrudes out of the SWNCT. This structural change increases the interaction between the twofold coordinated carbon and its surrounding atoms, and, in turn, causes the deep levels broaden and the local energy gap shrunken.

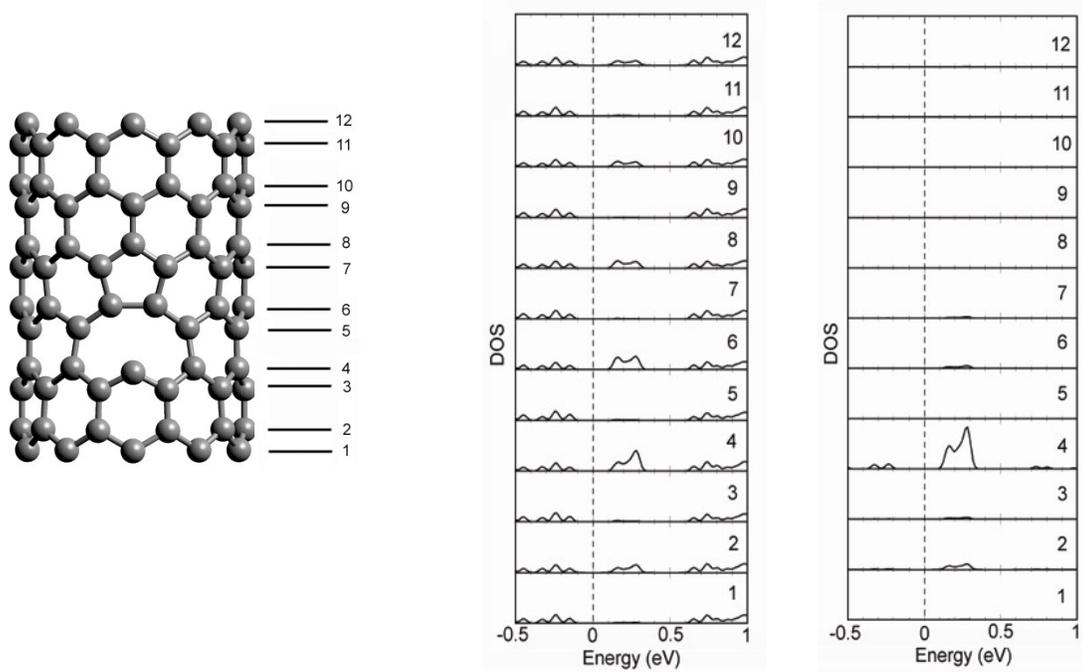

Figure 5. LPDOS of a (10,0) SWCNT with one vacancy out of 119 carbon atoms. (a) indicates the definition of the layers with the twofold coordinated carbon in the fourth layer and the newly formed C-C bond in the sixth layer. (b) is the LPDOS for the Px and Py orbitals of the carbon atoms in the particular layer, where the dash line is the Fermi level. (c) is the LPDOS for the Pz orbitals of the carbon atoms in the layer.

## Conclusions

This study provides insight into the influence of an isolated single vacancy defect in zigzag and armchair SWCNTs. The relationship between the local energy gap and the vacancy density has been found from the first principles. The bandstructure of an armchair SWCNT is very sensitive to the vacancy density which influences almost the entire bandstructure. However, no simple correlation between the vacancy density and bandgap has been determined.The impact range of a vacancy is characterized by structural deformation analysis made within the classical molecular dynamics framework as well. These results let us relate the structural deformation to the local energy gap variation. The strain increases with the defect density, causing the twofold coordinated carbon to protrude outward and, ultimately, reducing the local energy gap. This explanation interprets all data coherently.

## Acknowledgments

We acknowledge the financial support from the Bulgarian Ministry of education and science under the contract No. WU F05/06 and a support from the Project-based Personnel Exchange Programme between the NSC and BAS (2006).